\documentclass[reprint,superscriptaddress]{revtex4-1}
\usepackage{amsmath,amssymb,graphicx,float}
\usepackage[euler]{textgreek}
\begin{document}
\title{Nonlinear polarization dynamics of Kerr beam self-cleaning in a GRIN multimode optical fiber}

\author{Katarzyna Krupa}\email[]{Corresponding author: katarzyna.krupa@unibs.it}
\affiliation{Dipartimento di Ingegneria dell'Informazione, Universit\`a di Brescia, via Branze 38, 25123, Brescia, Italy}

\author{Graciela Garmendia Casta\~ neda}
\affiliation{Dipartimento di Ingegneria dell'Informazione, Universit\`a di Brescia, via Branze 38, 25123, Brescia, Italy}

\author{Alessandro Tonello}
\affiliation{Universit\'e de Limoges, XLIM, UMR CNRS 7252, 123 Av. A. Thomas, 87060 Limoges, France}

\author{Alioune Niang}
\affiliation{Dipartimento di Ingegneria dell'Informazione, Universit\`a di Brescia, via Branze 38, 25123, Brescia, Italy}

\author{Denis S. Kharenko}
\affiliation{Novosibirsk State University, 1 Pirogova street, 630090 Novosibirsk, Russia}

\author{Marc Fabert}
\affiliation{Universit\'e de Limoges, XLIM, UMR CNRS 7252, 123 Av. A. Thomas, 87060 Limoges, France}

\author{Vincent Couderc}
\affiliation{Universit\'e de Limoges, XLIM, UMR CNRS 7252, 123 Av. A. Thomas, 87060 Limoges, France}

\author{Guy Millot}
\affiliation{Universit\'e Bourgogne Franche-Comt\'e, ICB UMR CNRS 6303, 9 Av. A. Savary, 21078 Dijon, France}

\author{Umberto Minoni}
\affiliation{Dipartimento di Ingegneria dell'Informazione, Universit\`a di Brescia, via Branze 38, 25123, Brescia, Italy}

\author{Daniele Modotto}
\affiliation{Dipartimento di Ingegneria dell'Informazione, Universit\`a di Brescia, via Branze 38, 25123, Brescia, Italy}

\author{Stefan Wabnitz}
\affiliation{Dipartimento di Ingegneria dell'Informazione, Universit\`a di Brescia, via Branze 38, 25123, Brescia, Italy}
\affiliation{Novosibirsk State University, 1 Pirogova street, 630090 Novosibirsk, Russia}
\affiliation{Istituto Nazionale di Ottica del Consiglio Nazionale delle Ricerche (INO-CNR), via Branze 45, 25123 Brescia, Italy}

%% To be edited by editor
% \dates{Compiled \today}

%\ociscodes{(190.4370) Nonlinear optics, fibers; (190.3270) Kerr effect; (190.5940) Self-action effects.}

%% To be edited by editor
% \doi{\url{http://dx.doi.org/10.1364/XX.XX.XXXXXX}}

\begin{abstract}
We experimentally study polarization dynamics of Kerr beam self-cleaning in a graded-index multimode optical fiber. We show that spatial beam cleaning is accompanied by nonlinear polarization rotation, and a substantial increase of the degree of linear polarization.  
\end{abstract}

%\setboolean{displaycopyright}{true}

\maketitle

%\section{Introduction}

Research on multimode optical fibers (MMFs) has attracted a strong and renewed attention over the last few years. The first experimental observations of various nonlinear effects have been recently reported, including, e.g., multimode optical solitons \cite{Wise:0,Wright:1}, geometric parametric instability \cite{Krupa:2}, spatiotemporal mode-locking \cite{Wright:3}, and Kerr-induced beam self-cleaning.  The latter was demonstrated in graded-index multimode fibers (GRIN MMFs)  \cite{Krupa:4,Wright:4,Liu}, in active Yb-doped MMFs with quasi step-index profile  in a cavityless \cite{Guenard:5} and in a cavity configuration \cite{Guenard:6,Wright:3}, and in a multimode photonic crystal fiber \cite{Dupiol:6}. Nonlinear mode mixing has been also found to be at the origin of the strong temporal reshaping and pulse shortening which accompany self-cleaning in GRIN MMFs in the quasi-continuous wave regime \cite{Krupa:7}.

In nonpolarization maintaining optical fibers, the polarization components of guided modes are linearly coupled, owing to the unavoidable presence of weak random birefringence, that is induced by manufacturing imperfections during the fiber drawing process, and by externally applied stress or bending. As a consequence, multimode fibers do not generally maintain the input state of polarization (SOP). The evolution of the degree of polarization in MMFs was theoretically and experimentally studied in the linear propagation regime  \cite{Zhan:8, Yu:9, Kizevetter:10, Steeger:11}. For instance, Refs.\cite{Zhan:8, Steeger:11} experimentally showed that although a relatively short piece of step-index MMF may be able to preserve its input SOP, the degree of linear polarization (DOLP) of light decreases exponentially with increasing fiber length. 
%This fact in turn leads to a depolarization of a polarized beam launched into a fiber. Depolarization effects in multimode fibers have been theoretically and experimentally studied in the past in a linear regime of beam propagation. 
Recent experiments have shown that spatial wavefront shaping of the input beam may permit the control of the SOP emerging from MMFs \cite{Xiong:12}. 
Models for weak and strong mode coupling in nonlinear multimode fibers, exhibiting rapidly varying birefringence, were studied in Ref.\cite{Mumtaz}

%The objective of this paper is instead to point out 
In this letter, we experimentally address the polarization properties of nonlinear multimode beam propagation. In particular, we consider the polarization dynamics of spatial beam self-cleaning, which is induced in GRIN MMFs by the Kerr effect. We focus our attention on the nonlinear evolution of input beams with a linear SOP, and reveal that the onset of Kerr self-cleaning is accompanied by a significant increase of the DOLP at the fiber output. Moreover, we show that the self-cleaned beam experiences nonlinear polarization rotation, which eventually decreases the time-averaged DOLP of a pulsed beam at high powers. 
%The combination of high beam quality with high degree of beam polarization

%\section{Experimental set-up}
%\label{sec:examples}

In our experiments, we used standard, commercially available, 52/125 GRIN MMFs, that we pumped by using a Nd:YAG microchip laser delivering 500 ps pulses at 1064 nm, with the repetition rate of 500 Hz. We focused a laser beam at the input face of the fibers, with a full width of half maximum in intensity (FWHMI) diameter of 25 $\mu$m. In a last experiment, we also studied the influence of the input beam diameter on the output nonlinear polarization properties.%, and 81\% of coupling efficiency, 
This enables the controlled excitation of high-order guided modes \cite{Krupa:4}. %($\sim$ 120 modes in both polarizations). 
The input beam had a Gaussian spatial shape, and a linear SOP. The GRIN MMF was loosely coiled on the table, forming rings of $\sim$ 20 cm of diameter. We used a 3-axis translation stage to control the initial coupling conditions: we adjusted the position of the launched beam to be at the center of the fiber core, while maximizing the coupling efficiency up to 80\%. We used an optical spectrum analyzer (OSA) and a CCD camera to analyze the spectral and spatial properties of the output beam. 

In order to study the polarization dynamics of nonlinear multimode interactions and its connection to Kerr beam self-cleaning, we implemented a classical method used originally by Stokes in 1852. Namely, we measured at the fiber output the Stokes parameters of light by using a quarter wave-plate, followed by a rotating linear polarizer and a power meter \cite{Goldstein:13}. The Stokes parameters were averaged in time and in space over the whole output beam, owing to the long response time and the wide active area of the power meter.
To further characterize the polarization state of the output beam, we calculated the total degree of polarization $DOP = {\sqrt{ S_1^2 + S_2^2 + S_3^2}}/{S_0}$, the degree of linear polarization $DOLP = {\sqrt{ S_1^2 + S_2^2}}/{S_0}$, the degree of circular polarization $DOCP = {S_3}/{S_0}$, and the azimuth of linear polarization $\tan(2\psi) = {S_2}/{S_1}$, based on the previously determined Stokes parameters $S_0$, $S_1$, $S_2$ and $S_3$.

%To study the polarization dynamics of nonlinear multimode interactions and its connection to Kerr beam self-cleaning phenomenon, we implemented a classical method used originally by Stokes in 1852; we placed, at the fiber output, a quarter wave-plate followed by rotating linear polarizer and power meter, and we determined Stokes parameters by using the following relations: 

%\begin{subequations}
%\label{eq:S0}
%\begin{equation}
%%\label{eq:S0}
%S_0=P_{out}(90^\circ,0^\circ) + P_{out}(0^\circ,0^\circ)
%\end{equation}
%\begin{equation}
%%\label{eq:S1}
%S_1=P_{out}(90^\circ,0^\circ) - P_{out}(0^\circ,0^\circ)
%\end{equation}
%\begin{equation}
%%\label{eq:S2}
%S_2=P_{out}(45^\circ,0^\circ) - P_{out}(135^\circ,0^\circ)
%\end{equation}
%\begin{equation}
%%\label{eq:S3}
%S_3=P_{out}(45^\circ,90^\circ) - P_{out}(135^\circ,90^\circ)
%\end{equation}
%\end{subequations}	

%where $P_{out}(\theta,\phi)$ is a function of an angle of the polarizer transmission axis ($\theta$) with respect to the horizontal plane, and a phase delay between the light components of the slow and fast axes of the wave-plate ($\phi$). The values of Stokes parameters were measured on average in time and in space over whole beam. To further characterize the polarization state of the beam at the fiber output we calculated the following parameters based on the previously determined Stokes parameters:

\begin{figure}
\centering
\centerline{\includegraphics[width=\linewidth]{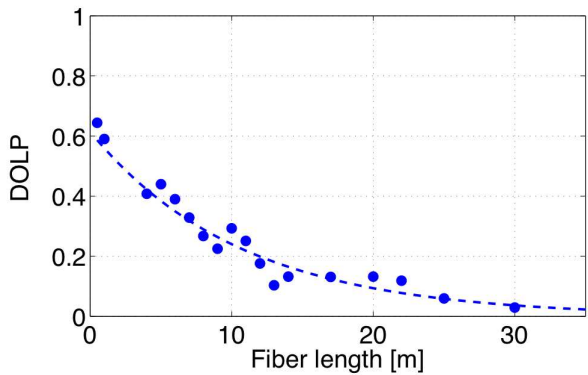}}
\caption{Measured DOLP (degree of linear polarization) as a function of fiber length under quasi-linear propagation conditions, i.e., with $P_{\text{in}}$ = 0.4 kW. Dashed line: exponential fit. Minimum length of the measured fiber: 0.5 m}
\label{fig:Fig1}
\end{figure}

%\section{Experimental results}

%Interestingly, we noticed that when rotating the linear polarization state of the input laser beam, the direction of the polarization state of the output beam follows these changes, and it also rotates by nearly the same angle, despite the fact that our multimode fiber is intrinsically randomly birefringent. 
%with up to a four-fold increase of its average degree of polarization, and a dominant linear polarization state.

We started our investigations by measuring the depolarization (of the linearly polarized input beams) along the length of the GRIN MMF in the low power regime ($P_{\text{in}}$$~$=$~$0.4 kW), by using the cut-back method. The results are reported in Fig.\ref{fig:Fig1}, which illustrates the exponential degradation of the DOLP as a function of fiber length, owing to random polarization coupling. The observed exponential decrease of DOLP with distance is similar to what previously experimentally reported for step-index MMFs \cite{Zhan:8, Yu:9}. Fig.\ref{fig:Fig1} shows that the DOLP drops from about 0.65 to below 0.2 for a GRIN fiber length of 11 m. 

\begin{figure}[t]
\centering
\centerline{\includegraphics[width=\linewidth]{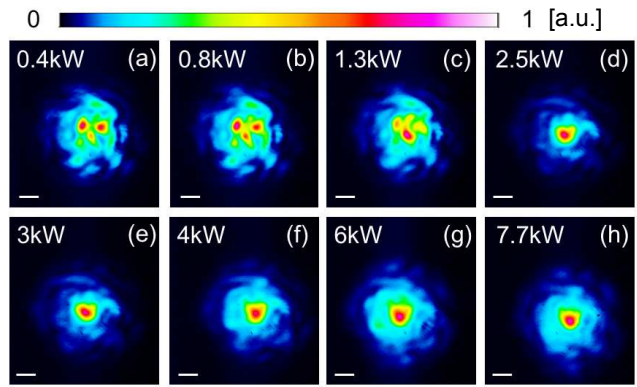}}
\caption{Near-field images of fiber output beams as a function of input peak power $P_{\text{in}}$ (a-h), showing the Kerr-induced beam self-cleaning effect. Fiber length: 11 m; Scale bars: 10 $\mu$m}
\label{fig:Fig2}
\end{figure}

Next, we studied the nonlinear dynamics of the SOP and DOP in the regime of Kerr-induced beam cleaning, by using another 11-m long span of similar GRIN MMF. As can be seen in Fig.\ref{fig:Fig2}, for input peak powers above a threshold value of about 2.5 kW, an initially randomly speckled output beam evolves into a robust bell-shaped profile, whose cross section turns out to be close to the fundamental mode of the GRIN fiber. At first, we placed a simple rotating polarizer at the fiber output, and we investigated the output spatial beam profile as a function of the angle of the polarizer transmission axis, for input peak powers below and above the power threshold for Kerr beam self-cleaning, respectively. 

\begin{figure*}
\centering
\centerline{\includegraphics[width=\linewidth]{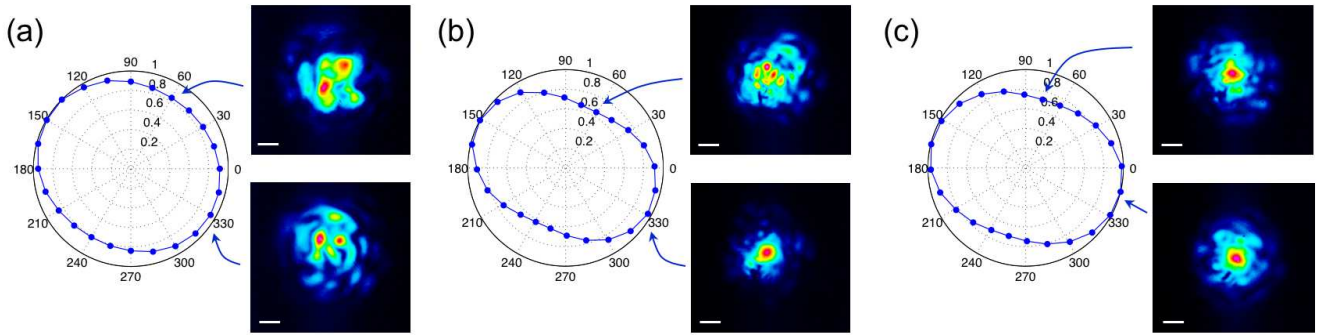}}
\caption{Normalized power for one output polarizer revolution for input peak power (a) $P_{\text{in}}$ = 0.4 kW (linear regime), (b) $P_{\text{in}}$ = 3 kW, (c) $P_{\text{in}}$$~$=$~$6\,kW (two examples in Kerr beam self-cleaning regime). Insets: corresponding spatial patterns measured at the output face of the fiber along two main perpendicular polarization axes. Fiber length: 11 m. Input polarization: DOLP=1. Scale bars: 10 $\mu$m}
\label{fig:Fig3}
\end{figure*}

Figure \ref{fig:Fig3} illustrates the evolution of the output power $P_{\text{out}}$ at 1064 nm normalized to its maximum value, corresponding to a complete rotation of the axis of the output polarizer. In Fig. \ref{fig:Fig3} we compare the low power case of $P_{in}$ = 0.4 kW (Fig.\ref{fig:Fig3}(a)), with the situation just above ($P_{in}$ = 3$kW$) and more than twice above ($P_{in}$$~$=$~$6$kW$) the self-cleaning threshold (Fig.\ref{fig:Fig3}(b) and Fig.\ref{fig:Fig3}(c)), respectively. Insets of panels (a)-(c) of Fig.\ref{fig:Fig3} show the corresponding near-field images of the output beams, for two selected output polarizer orientations, corresponding to orthogonal polarization axes of the output light. 

The corresponding near-field output patterns, i.e., for the same input beam but when the output polarizer is removed, are presented in panels (a), (e), and (g) of Fig.\ref{fig:Fig2}. As we can see from \ref{fig:Fig3}(a), at the relatively low power of 0.4 kW, light at the fiber output exhibits a speckled spatial pattern, for both perpendicular angles of polarized transmission axes. We verified that the same occurs for all the other directions of polarization. Fig.\ref{fig:Fig3}(b) shows that near the self-cleaning threshold power of 3 kW, a bell-shaped output spatial profile (see also Fig.\ref{fig:Fig2}(e)) is only observed around the main polarization axis. Whereas a speckled beam still remains along the orthogonal axis. Nevertheless, when the input peak power grows well above the self-cleaning threshold, a bell-shaped beam profile is now visible for the two main perpendicular directions. Again, we verified that the same occurs for a complete rotation of the output polarizer transmission axis. This is because the power threshold for Kerr beam cleaning is reached for all polarization directions. An example measured at $P_{\text{in}}$ $=$ 6 kW is illustrated in the panel (c) of Fig.\ref{fig:Fig3}.

\begin{figure}[!hb]
\centering
\centerline{\includegraphics[width=\linewidth]{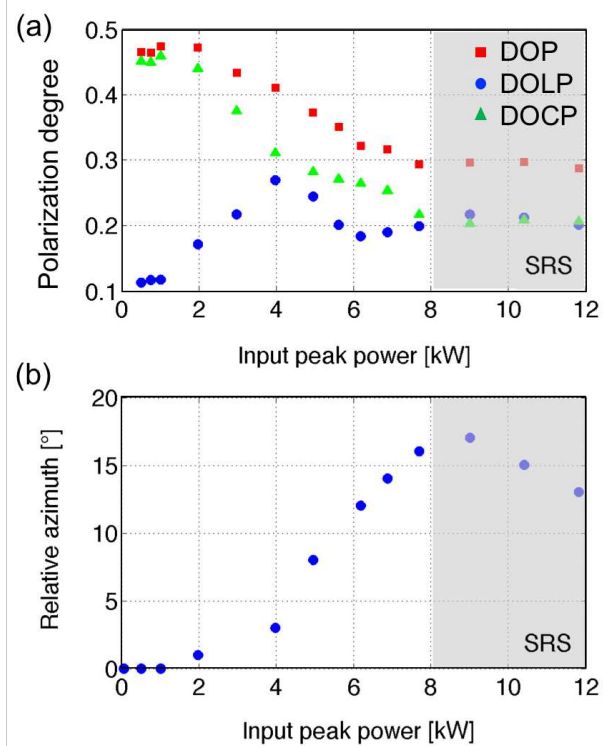}}
\caption{Polarization dynamics of beam propagation in a GRIN MMF. (a) Measured DOLP (blue dots), DOP (red squares) and DOCP (green triangles) as a function of input peak power ($P_{\text{in}}$); (b) Relative azimuth of linear polarization as a function of $P_{\text{in}}$; Gray area: regime of SRS generation.}
\label{fig:Fig4}
\end{figure}

Let us consider next the effect of Kerr beam self-cleaning on the output DOLP. As we reported in Fig.\ref{fig:Fig1}, owing to random polarization mode coupling a linearly polarized laser beam has a DOLP that decreases gradually with the propagation distance in a GRIN MMF. Fig.\ref{fig:Fig4}(a) (blue dots) shows an example of the evolution of the output DOLP as a function of input peak power, for the 11-m long GRIN fiber span. Note that in this fiber span the linear depolarization effects are stronger than in the fiber used for the cut-back measurements reported in Fig.\ref{fig:Fig1}, since after 11-m the DOLP drops down to 0.1 at low powers. Fig.\ref{fig:Fig4}(a) (blue dots)
shows that the linear degree of polarization first rapidly increases up to DOLP=0.26, when the input power crosses the power threshold for Kerr self-cleaning and reaches $P_{in}$ = 4 kW. Next, the DOLP decreases when the beam power grows larger past 4 kW. 

Because the size of the fundamental mode (7.45 $\mu$m of FWHMI) is much smaller than the core diameter of the GRIN MMF (52 $\mu$m), it is reasonable to think that the SOP of light carried by the fundamental mode is relatively insensitive to technological irregularities and external disturbances along the fiber. Therefore the SOP of the fundamental mode of a GRIN MMF is more robust than the SOP in nonpolarization-preserving single-mode fibers. Energy transfer towards the fundamental mode owing to Kerr beam self-cleaning may thus explain the initial re-polarization, and $\sim$ 2.5-fold increase of DOLP observed in Fig.\ref{fig:Fig4}(a) (blue dots). The subsequent decrease of DOLP which is observed at higher input powers is likely to result instead from nonlinear polarization rotation, which becomes important for $P_{in}$ above 4 kW, as illustrated by Fig.\ref{fig:Fig4}(b). Note that since we used a pulsed beam, the SOP varies along the temporal profile of the pulse with different rotation values, while our SOP measurements are averaged in time. This temporal averaging results in a nonlinear depolarization effect.

In panel (b) of Fig.\ref{fig:Fig4} we illustrate the observed shift of the azimuth $\psi$ of the main linear polarization axis of the output beam as a function of input power. We may clearly notice in Fig.\ref{fig:Fig4}(b) the change of more than 15 degrees of the relative azimuth $\psi$. This demonstrates that output polarization rotation occurs when the input power increases. Such nonlinear polarization rotation reaches a maximum at the threshold for stimulated Raman scattering (SRS). 

In order to fully characterize the SOP of output light, in panel (a) of Fig.\ref{fig:Fig4} we also present the experimental measurements of the output DOCP and DOP versus input power (green triangles and red squares, respectively). It can be observed that both the DOCP and the DOP decrease with power, an effect which can be ascribed to the nonlinear polarization rotation observed in Fig.\ref{fig:Fig4}(b), combined with temporal averaging across the pulse profile: different time sections rotate their output SOP by different amounts, such that the average DOP (DOCP) is decreased by fiber nonlinearity. Note that an increase of input power also results in pulse progressive pulse narrowing \cite{Krupa:7}.

\begin{figure}[t]
\centering
\centerline{\includegraphics[width=\linewidth]{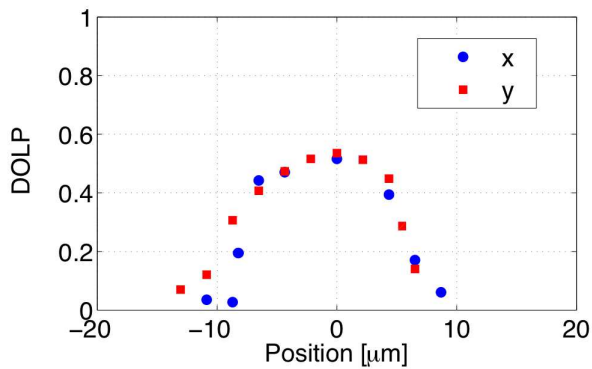}}
\caption{Measured DOLP as a function of transverse position across the beam profile along x-axis (y=0) (blue dots) and y-axis (x=0) (red squares), and at $P_{\text{in}}$ = 4 kW.}
\label{fig:Fig5}
\end{figure}

Since the self-cleaned beam only occupies a relatively small area in the transverse dimension of the output beam, it is important to analyze the spatial distribution of the output DOLP. For this purpose, we filtered the 46 times magnified output near field by a diaphragm with an opening of about $\sim$ 600\,$\mu$m in diameter, which corresponds to a disc area of $\sim$ 13 $\mu$m diameter on the fiber output face. This is about twice the size of the self-cleaned, bell-shaped central beam.
The diaphragm was then moved across the output beam horizontally at constant latitude, as well as vertically at constant longitude. %We shifted along the horizontal coordinate (vertical coordinate) the relative position of this diaphragm with respect to the beam image, while keeping the vertical (horizontal) position fixed. 
The results are presented in Fig.\ref{fig:Fig5} by using blue dots for shifts along the x-axis (y=0), or red squares for shifts along the y-axis (x=0). The input power was kept fixed at $P_{\text{in}}$ = 4 kW, which is the value leading to a maximum DOLP when analyzing the whole beam (see Fig.\ref{fig:Fig4}(a) (blue dots). As can be seen in Fig.\ref{fig:Fig5}, at the beam center the DOLP increases above 0.6, indicating a significant degree of re-polarization for the self-cleaned bell-shaped beam.

\begin{figure}[H]
\centering
\centerline{\includegraphics[width=\linewidth]{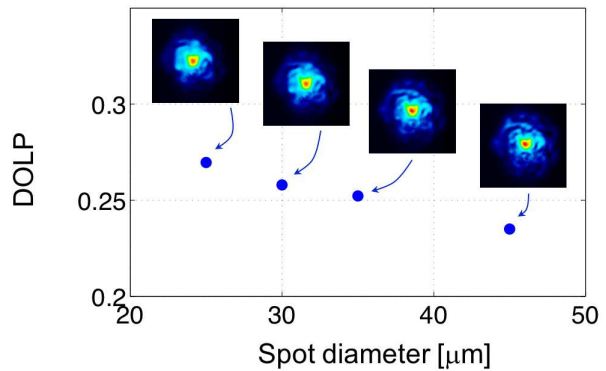}}
\caption{Measured DOLP of the output beam for various FWHMI diameters of the spot focused at the input face of the fiber at fixed output power of 3.2$~$kW ($i.e.$ at $P_{\text{in}}$ = 4\,kW for the diameter of 25 $\textmu$m). Insets: corresponding near-field images of the fiber output beams.}
\label{fig:Fig6}
\end{figure}

In another experiment, we investigated the influence of the beam diameter at the input of the fiber on the polarization dynamics in the regime of Kerr beam self-cleaning. Since the coupling efficiency decreases with the beam diameter, we fixed the output power at 3.2 kW, which corresponds to $P_{\text{in}}$ = 4 kW for the input beam diameter of 25 $\mu$m. The results of DOLP measurements are shown in Fig.\ref{fig:Fig6}. 
Here insets illustrate the corresponding near-field images. As can be seen, the output DOLP decreases when the spot diameter grows larger, which may be understood as follows. Focusing the beam at the fiber input face with a larger diameter leads to an energy distribution over a larger number of modes, so that the input fundamental mode energy is reduced. As a consequence, self-cleaning induced energy transfer towards the bell-shaped beam, and the associated nonlinear re-polarization, are both less efficient, as demonstrated by Fig.\ref{fig:Fig6}.

%\section{Numerical simulations}

%?????????????????????????????

%\section{Conclusions}

To conclude, we investigated nonlinear polarization dynamics in the regime of beam self-cleaning in nonlinear GRIN MMFs. We experimentally demonstrated that complex nonlinear mode mixing leading to Kerr-induced beam self-cleaning is accompanied by a nonlinear re-polarization effect. 
%Kerr-induced beam self-cleaning process which reshapes the transverse profile, may also impact the polarization state. 

A linearly polarized input pump beam, which is almost depolarized when measured at the fiber output in the linear propagation regime, is partially re-polarized at powers close to the threshold for self-cleaning. Moreover, we also observed the evidence of nonlinear polarization rotation of the self-cleaned beams.

This effect could be used to implement an ultrafast saturable absorber mechanism for mode-locked lasers based on multimode fibers \cite{Mafi:14, Wang:15}. These effects are relevant since the polarization properties of laser beam delivery systems are crucial for many applications. Fully polarized high quality beams are essential, for example, for applications in coherent communications and medical tissue diagnosis. Whereas, partially polarized laser beams are required in laser micro-machining.

\section*{Funding Information}

%We acknowledge the financial support from: 
The European Research Council (ERC) under the European Union's Horizon 2020 research and innovation programme (grant No. 740355); French "Investissements d'Avenir" programme under the ISITE-BFC 299 project (ANR-15-IDEX-0003); ANR Labex ACTION (ANR-11-LABX-0001-01); iXcore research foundation. K. K. has received funding from the European Union's Horizon 2020 research and innovation programme under the Marie Sklodowska-Curie grant agreement No. 713694 (MULTIPLY). D. Kh. was supported by the Russian Ministry of Science and Education (Grant 14.Y26.31.0017). 

\section*{Acknowledgment} We thank G. Brambilla and A. Barth\'el\'emy for helpful remarks.

%\section*{References}

% Bibliography
%\bibliography{sample}

\end{document}